\documentstyle[a4,12pt]{article}
\begin{document}
\pagestyle{empty}
\begin{titlepage}
\vspace{2.0 truecm}
\begin{center}
\begin{Large}
{\bf Model independent analysis of the trilinear } \\ [0.4cm]
{\bf  gauge boson couplings at LC: role of} \\ [0.4cm] 
{\bf polarized cross sections}
\hskip 2pt\footnote{Contribution to the Workshop on ``$e^+e^-$ collisions at
$TeV$ energies: The physics potential'', DESY 97-123E} 
\end{Large}

\vspace{2.0cm}
             
{\large  A. A. Pankov\hskip 2pt\footnote{Permanent address: Gomel
Polytechnical Institute, Gomel, 246746 Belarus.\\
E-mail PANKOV@GPI.GOMEL.BY}
}\\[0.3cm]
International Centre for Theoretical Physics, Trieste, Italy\\
Istituto Nazionale di Fisica Nucleare, Sezione di Trieste, 34127 Trieste,
Italy\\
\vspace{5mm}

{\large  N. Paver\hskip 2pt\footnote{Also supported by the Italian
Ministry of University, Scientific Research and Technology (MURST).}
}\\[0.3cm]
Dipartimento di Fisica Teorica, Universit\`{a} di Trieste, 34100
Trieste, Italy\\
Istituto Nazionale di Fisica Nucleare, Sezione di Trieste, 34127 Trieste,
Italy
\end{center}

\vspace{2.0cm}

\begin{abstract}
\noindent
By means of a model-independent analysis, we discuss the constraints on 
anomalous trilinear gauge-boson couplings that can be obtained from the study 
of electron-positron annihilation into $W$ pairs at LC with 
$\sqrt{s}=0.5\ TeV$ and $1\ TeV$. We consider the general $CP$ conserving 
anomalous effective Lagrangian, as well as some specific models with reduced 
number of independent couplings. The analysis is based on combinations of 
observables with initial and final state polarizations, that allow to 
separately constrain the different couplings and to improve the corresponding 
numerical bounds.\par
      
\vspace*{3.0mm}

\noindent
\end{abstract}
\end{titlepage}
\pagestyle{plain}
\setlength{\baselineskip}{1.3\baselineskip}

The values of the $WW\gamma$ and $WWZ$ couplings, and the corresponding 
non abelian gauge structure of the Standard Model (SM), have not been
tested yet. In this regard, the reaction   
\begin{equation}e^++e^-\to W^++W^-\label{proc}\end{equation}
at high energy $e^+e^-$ colliders is particularly important  
because, for such process, deviations from the SM predictions due to anomalous 
values of the trilinear coupling constants are significantly enhanced 
by increasing the CM energy, and the related sensitivity is improved.
The general anomalous trilinear gauge boson Lagrangian has a complicated 
structure, containing both CP violating and CP conserving interactions. 
The set of cross-section measurements for process (\ref{proc}), relevant to 
the CP violating couplings and their separation, was discussed in 
Ref.\cite{gounaris1}.\footnote{A discussion of constraints on  
the C and P violating (but CP conserving) anapole coupling from measurements 
of process (\ref{proc}) with longitudinally polarised leptons was given in 
Ref.\cite{anom2}.} 
\par 
In what follows, we examine the possibility of constraining the CP conserving 
effective Lagrangian, which can be expressed as follows: 
\cite{hagiwara,gounaris2}
\begin{eqnarray}{\cal L}&=&-ie\hskip 2pt\left[A_\mu\left(W^{-\mu\nu}W^+_\nu-
W^{+\mu\nu}W^-_\nu\right)+F_{\mu\nu}W^{+\mu}W^{-\nu}\right]-ie\hskip 2pt 
x_{\gamma}\hskip 2pt F_{\mu\nu}W^{+\mu}W^{-\nu}
\nonumber\\ 
&-& ie\hskip 2pt \left(\cot\theta_W+\delta_Z\right)\hskip 2pt \left[Z_\mu
\left(W^{-\mu\nu}W^+_\nu-W^{+\mu\nu}W^-_\nu\right)+Z_{\mu\nu}W^{+\mu}W^{-\nu}
\right]\nonumber\\ 
&-&ie\hskip 2pt x_Z\hskip 2pt Z_{\mu\nu}W^{+\mu}W^{-\nu}
 +ie\hskip 2pt\frac{y_{\gamma}}{M_W^2}\hskip 2pt F^{\nu\lambda}W^-_{\lambda\mu}
W^{+\mu}_{\ \ \nu}+ie\hskip 2pt\frac{y_Z}{M_W^2}\hskip 2pt 
Z^{\nu\lambda}
W^-_{\lambda\mu}W^{+\mu}_{\ \ \nu}\hskip 2pt ,\label{lagra}\end{eqnarray}
where $W_{\mu\nu}^{\pm}=\partial_{\mu}W_{\nu}^{\pm}-
\partial_{\nu}W_{\mu}^{\pm}$ and ${Z_{\mu\nu}=\partial_{\mu}Z_{\nu}-
\partial_{\nu}Z_{\mu}}$. 
Clearly, in the SM: $\delta_Z=x_{\gamma}=x_Z=y_{\gamma}=y_Z=0$. 
As Eq.(\ref{lagra}) shows, in general we have five independent couplings. 
\par 
Models with smaller number of anomalous couplings naturally obtain in the 
framework of the effective theory, where the existence of some new interaction,  
acting at a mass scale $\Lambda$ much higher than the Fermi scale, is assumed. 
In this case, anomalous couplings originate as remnants of such interaction 
at lower energy scales, in the form of corrections to the SM suppressed by
inverse powers of $\Lambda$ \cite{zeppenfeld1}. Specifically, the effective 
weak interaction Lagrangian is expanded as:
\begin{equation}{\cal L}_W={\cal L}_{SM}+\sum_d\sum_k
\frac{f^{(d)}_k}{\Lambda^{d-4}}{\cal O}^{(d)}_k, \label{lagra1}\end{equation}
where ${\cal L}_{SM}$ is the SM interaction and the second term, representing
the `low-energy' new interaction effects, is expressed in terms of 
$SU(2)\times U(1)$ gauge invariant operators ${\cal O}^{(d)}_k$
with dimension $d$ made of $\gamma$, $W$, $Z$ and Higgs 
fields,\footnote{Such spontaneously broken gauge invariance requirement 
for the new interaction is 
naturally justified on phenomenological grounds. Similarly, 
lepton couplings are assumed to be unaffected by the new interaction.} 
times their respective coupling constants $f^{(d)}_k$ not fixed by the symmetry.  
Truncation of the sum in Eq.~(\ref{lagra1}) to the lowest
significant dimension, $d=6$, limits the number of allowed independent
operators (and their corresponding constants) to three
\cite{zeppenfeld1}-\cite{buch}:
\begin{eqnarray}{\cal O}^{(6)}_{WWW}&=&Tr\left[{\hat W}_{\mu\nu}
{\hat W}^{\nu\rho}{\hat W}^{\mu}_{\rho}\right],\nonumber\\
{\cal O}^{(6)}_W&=&\left(D_\mu\Phi\right)^\dagger{\hat W}^{\mu\nu}
\left(D_\nu\Phi\right),\nonumber\\
{\cal O}^{(6)}_B&=&\left(D_\mu\Phi\right)^\dagger{\hat B}^{\mu\nu}
\left(D_\nu\Phi\right),\label{oper}\end{eqnarray}
where $\Phi$ is the Higgs doublet and, in terms of the $B$ and $W$
field strengths: ${\hat B}^{\mu\nu}=i(g^\prime/2)B^{\mu\nu}$,
${\hat W}^{\mu\nu}=i(g/2){\vec\tau}\cdot{\vec W}^{\mu\nu}$ with $\vec\tau$
the Pauli matrices. Eq.~(\ref{oper}) implies the relations among the anomalous coupling 
constants:
\begin{equation}x_\gamma=\cos^2\theta_W\hskip 2pt
\left(f^{(6)}_B+f^{(6)}_W\right)\hskip 2pt\frac{M^2_Z}{2\Lambda^2};
\qquad y_\gamma=f^{(6)}_{WWW}
\hskip 2pt\frac{3M^2_Wg^2}{2\Lambda^2}; \label{deltaz}\end{equation}
\begin{equation}\delta_Z=\cot\theta_W\hskip 2pt f^{(6)}_W
\hskip 2pt\frac{M^2_Z}{2\Lambda^2};
\quad x_Z=-\tan\theta_W\hskip 2pt x_\gamma;
\quad y_Z=\cot\theta_W\hskip 2pt y_\gamma,\label{xgamma}\end{equation}
so that only three (e.g., $x_\gamma$, $y_\gamma$ and $\delta_Z$) are 
independent. 
\par
More relations occur from further specializations. An example is represented 
by the two-parameter `HISZ scenario' \cite{zeppenfeld1}, that assumes the 
relation $f^{(6)}_B=f^{(6)}_W$ in the above equations, such that
\begin{equation}
\delta_Z=\frac{1}{2\sin\theta_W\cos\theta_W}\hskip 2pt x_\gamma,
\quad x_Z=-\tan\theta_W\hskip 2pt x_\gamma,
\quad y_Z=\cot\theta_W\hskip 2pt y_\gamma.
\label{hisz}\end{equation}
\par 
As an alternative to the above formulation, the number of independent 
trilinear anomalous can be reduced by the assumption of
global $SU(2)_L$ symmetry 
of Eq.~(\ref{lagra}), which directly implies the relation
$x_Z=-\tan\theta_W\hskip 2pt x_\gamma$,\footnote{The same as in 
Eq.~(\ref{xgamma}).} plus the neglect dimension 6 quadrupole
operators leading to $y_\gamma=y_Z=0$, and imposing the cancellation 
at order $s^2$ of the tree-level unitarity violating contributions to 
$WW$ scattering, which in turn implies the relation 
$\delta_Z=x_\gamma/\sin\theta_W\hskip 1pt\cos\theta_W$
\cite{kuroda,bilchak}. In this case, therefore, only one independent 
parameter remains.
\par
As previously noticed, in the general case of Eq.~(\ref{lagra}) the five 
independent trilinear constants cannot be separately studied and constrained 
by the unpolarized cross section alone, which depends on all the couplings. 
Separate measurements of the cross sections for specific initial and final 
states polarizations, depending on independent combinations of the trilinear 
coupling constants, give the necessary additional information that allows to 
disentangle the couplings in a model independent way. Ideally, to that purpose,
the three 
the possible $W^+W^-$ polarizations ($LL$, $TL$ and $TT$), combined with the 
two longitudinal $e^-\hskip 2pt e^+$ ones ($RL$ and $LR$) should  
determine a sufficient set of observable cross sections.
\par 
The basic objects to be studied are the potential deviations of the 
polarized cross sections from the SM predictions due to finite values of 
anomalous couplings in Eq.~(\ref{lagra}):  

\begin{equation}\Delta\sigma=\sigma-\sigma_{SM}.\label{deltasig}\end{equation}
Limiting to the Born level $\gamma$-, $Z$- and $\nu$-exchange 
amplitudes:
\begin{eqnarray}d\sigma&\propto&
\vert{\cal A}(\gamma)+\Delta{\cal A}(\gamma)+
{\cal A}(Z)+\Delta{\cal A}(Z)+{\cal A}_1(\nu)\vert^2+
\vert{\cal A}_2(\nu)\vert^2,\nonumber \\
d\sigma_{SM}&\propto&
\vert {\cal A}(\gamma)+{\cal A}(Z)+{\cal A}_1(\nu)\vert^2
+\vert{\cal A}_2(\nu)\vert^2.\label{deltaa}\end{eqnarray}
In Eq.(\ref{deltaa}), it is convenient to distinguish the 
$\nu$- exchange amplitudes with $\vert\lambda-\bar\lambda\vert\leq 1$ from  
$\vert\lambda-\bar\lambda\vert=2$ ones, with $\lambda$ and $\bar\lambda$ 
the $W^-$ and $W^+$ helicities. Using the explicit 
helicity amplitudes given, {\it e.g.}, in Ref.\cite{gounaris2}, and the 
Lagrangian Eq.~(\ref{lagra}), the amplitudes deviations 
$\Delta{\cal A}$'s with initial beams and final $W$'s specific polarizations 
have the following dependence on the anomalous couplings:
\begin{eqnarray} \Delta{\cal A}_{LL}^{ab}(\gamma)&\propto&x_{\gamma}
\nonumber\\
\Delta{\cal A}_{LL}^{ab}(Z)&\propto&\left(x_Z+\delta_Z\hskip 1pt 
\frac{3-\beta_W^2}{2}\right)\hskip 1pt g_e^a,\label{deltall}\end{eqnarray}
\begin{eqnarray} \Delta{\cal A}_{TL}^{ab}(\gamma)&\propto&x_{\gamma}+y_{\gamma}
\nonumber\\
\Delta{\cal A}_{TL}^{ab}(Z)&\propto&\left(x_Z+y_Z+2\hskip 1pt\delta_Z\right)
\hskip 1pt g_e^a,\label{deltatl}\end{eqnarray}
and
\begin{eqnarray} \Delta{\cal A}_{TT}^{ab}(\gamma)&\propto&y_{\gamma}
\nonumber\\
\Delta{\cal A}_{TT}^{ab}(Z)&\propto&\left(y_Z+\delta_Z\hskip 1pt 
\frac{1-\beta_W^2}{2}\right)\hskip 1pt g_e^a.\label{deltatt}\end{eqnarray}
In Eqs.(\ref{deltall})-(\ref{deltatt}):  
${\beta_W=\sqrt{1-4M_W^2/s}}$, the indices $LL$, $TL$ and $TT$ refer to 
the final $W^-W^+$ polarizations $LL$, $TL+LT$ and $TT$ respectively, while 
the upper indices $a$ and $b$ refer to the $e^-$ $e^+$ $RL$ or $LR$ 
polarization. Furthermore, $g_e^R=\tan\theta_W$ and 
$g_e^L=g_e^R\left(1-1/2\sin^2\theta_W\right)$ represent the corresponding 
electron couplings. One can notice that $\sigma_{LL}$, $\sigma_{TL}$ and 
$\sigma_{TT}$ depend on the combinations 
($x_{\gamma},x_Z+\delta_Z(3-\beta_W^2)/2$), 
($x_{\gamma}+y_{\gamma},x_Z+y_Z+2\delta_Z$) and 
($y_{\gamma},y_Z+\delta_Z(1-\beta_W^2)/2$) respectively.
\par  
As a procedure to quantitatively assess the sensitivity of the different cross 
sections to the gauge boson couplings, we divide the experimentally significant range of the
production angle $\cos\theta$ (assumed to be $\vert\cos\theta\vert\leq 0.98$) 
into 10 `bins', and define the $\chi^2$ function:
\begin{equation}
\chi^{2}=\sum^{bins}_i\left[\frac{N_{SM}(i)-N_{anom}(i)}
{\delta N_{SM}(i)}\right]^2,\label{chi2}\end{equation}
where $N(i)= L_{int}\sigma_i\varepsilon_W$ represents the expected number of 
events in the $i$-th bin with $\sigma_i$  the corresponding cross section 
(either the SM or the anomalous one):
\begin{equation}
\sigma_i\equiv\sigma(z_i,z_{i+1})=
\int \limits_{z_i}^{z_{i+1}}\left({d\sigma}\over{dz}\right)dz,
\qquad\ z=\cos\theta.\label{sigmai}\end{equation}
The efficiency $\varepsilon_W$ for $W^+W^-$ reconstruction in the
various polarization states is taken as 
$\varepsilon_W\simeq 0.3$ \cite{frank}-\cite{anlauf} from the channel of
lepton pairs ($e\nu+\mu\nu$) plus two hadronic jets and the corresponding 
branching ratios. In fact, the actual value of $\varepsilon_W$ for polarized
final states might be considerably smaller, depending on experimental details
\cite{frank}, but definite estimates are not available at present. As a
compensation, for the time-integrated luminosity which is multiplied by 
$\varepsilon_W$ everywhere, we assume the rather conservative value (compared
with recent findings \cite{ruth}): $\displaystyle L_{int}=20\hskip 2pt fb^{-1}$ 
for the NLC(500) and $\displaystyle L_{int}=50\hskip 2pt fb^{-1}$ for the 
NLC (1000). 
\par 
In the case no deviations were observed in the cross sections under 
consideration, allowed regions for the anomalous coupling constants can be 
are obtained by adopting, as a criterion, that $\chi^2\leq\chi^2_{crit}$, 
where $\chi^2_{crit}$ is a number that corresponds to the chosen confidence 
level. Since each polarized cross section involves two well-defined 
combinations of anomalous couplings at a time, as 
Eqs.~(\ref{deltall})-(\ref{deltatt}) show, with two independent degrees of 
freedom 95\% CL bounds in each separate case are obtained by choosing 
$\chi^2_{crit}=6$.
\par
Since, in practice, initial beams polarization will not be perfect,  
to adapt the analysis to the possible experimental situation we should 
consider the cross section   
\begin{equation}
\frac{d\sigma}{dz}=\frac{1}{4}\left[(1+P_1)\cdot (1-P_2)\hskip 2pt
\frac{d\sigma^{RL}}{dz}
+(1-P_1)\cdot (1+P_2)\hskip 2pt
\frac{d\sigma^{LR}}{dz}\right],\label{gcr}
\end{equation}
where $P_1$ ($P_2$) are less than unity, and represent the actual degrees of 
longitudinal polarization of $e^-$ ($e^+$). Reasonable possibilities seems to 
be: $d\sigma^R/dz$ ($P_1=0.9$, $P_2=0$) and
$d\sigma^L/dz$ ($P_1=- 0.9$, $P_2=0$).
\par
We present numerical results case by case, starting from 
longitudinally polarized $e^-e^+\to W^-_LW^+_L$ production, for both
possibilities of the electron beam longitudinal polarization. 
The typical resulting area,
allowed to the combinations of anomalous couplings in Eq.~(\ref{deltall}) at
the $95\%$ CL for both $\sqrt s=0.5$ and $1\hskip 2pt TeV$,
can be directly read from Fig.~2 of Ref.{\cite{anom4}}. It implies the pair of
inequalities
\begin{equation}-\alpha_1^{LL}<x_{\gamma}<\alpha_2^{LL},\label{ll1}
\end{equation}
\begin{equation}-\beta_1^{LL}<x_Z+\delta_Z\frac{3-\beta_W^2}{2}<\beta_2^{LL},
\label{ll2}\end{equation}
so that only $x_\gamma$ is separately constrained at this stage.
Here, $\alpha_{1,2}^{LL}$ and $\beta_{1,2}^{LL}$ are the projections of the
combined allowed area on the horizontal and vertical axes.
\par
The same kind of analysis can be applied to the other polarized cross 
sections. From $e^+e^-\to W^+_TW^-_L+W^+_LW^-_T$ we obtain the allowed region
for the combinations of coupling constants in Eq.~(\ref{deltatl}), that 
implies the analogous inequalities:
\begin{equation}-\alpha_1^{TL}<x_{\gamma}+y_{\gamma}<\alpha_2^{TL},
\label{tl1}\end{equation}
\begin{equation}
-\beta_1^{TL}<x_Z+y_Z+2\delta_Z<\beta_2^{TL}.\label{tl2}\end{equation}
Finally, from $e^+e^-\to W^+_TW^-_T$ one obtains the corresponding
inequalities:
\begin{equation}-\alpha_1^{TT}<y_{\gamma}<\alpha_2^{TT},\label{tt1}
\end{equation}
\begin{equation}
-\beta_1^{TT}<y_Z+\frac{1-\beta_W^2}{2}\delta_Z<\beta_2^{TT}.
\label{tt2}\end{equation}
By combining Eqs.~(\ref{ll2})-(\ref{tt2}), one can very simply disentangle
the bounds for $\delta_Z$, $x_Z$ and $y_Z$:
\begin{equation}-\frac{1}{\beta_W^2}B_2<\delta_Z<\frac{1}{\beta_W^2}B_1,
\label{B1}\end{equation}
\begin{equation}-\left(\beta_1^{LL}+\frac{3-\beta_W^2}{2\beta_W^2}B_1\right)<
x_Z<\beta_2^{LL}
+\frac{3-\beta_W^2}{2\beta_W^2}B_2\label{B2},\end{equation}
\begin{equation}-\left(\beta_1^{TT}+\frac{1-\beta_W^2}{2\beta_W^2}B_1\right)
<y_Z<\beta_2^{TT}
+\frac{1-\beta_W^2}{2\beta_W^2}B_2,\label{B3}\end{equation}
where $B_1=\beta_1^{LL}+\beta_1^{TT}+\beta_2^{TL}$ and
$B_2=\beta_2^{LL}+\beta_2^{TT}+\beta_1^{TL}$. Adding these constraints to
those in Eqs.~(\ref{ll1}) and (\ref{tt1}) for $x_\gamma$ and $y_\gamma$,
we finally obtain, by this simple procedure, separate bounds for the 
five anomalous couplings. 
\par 
With the chosen inputs for the luminosity and the beam polarization quoted 
previously, numerical results are as reported in Tab.~\ref{tab:tab1}.
\begin{table}
\centering
\caption{ Model independent limits on the five $CP$ even nonstandard
gauge boson couplings at the $95\%$ CL.} \vskip 0.2cm
\begin{tabular}{|c|c|c|c|c|c|}
\hline
$\sqrt{s}\ (TeV)$ & $x_\gamma\hskip2pt (10^{-3})$ & $y_\gamma\hskip2pt
(10^{-3})$ & $\delta_Z\hskip2pt (10^{-3})$ & $x_Z\hskip2pt (10^{-3})$ &
$y_Z\hskip2pt (10^{-3})$
\\ \hline
$0.5$ & $-2.0\div 2.2$ & $-11.0\div 10.6$ &$-52\div 45$
& $-51\div 59$ & $-22\div 30$
\\ \hline
$1$ &$-0.6\div 0.6$ & $-3.2\div 3.4$ &$-19\div 16$
& $-18\div 20$ & $-5.7\div 6.2$
\\ \hline
\end{tabular}
\label{tab:tab1}
\end{table}
\par 
Tab.~\ref{tab:tab2} summarizes the numerical bounds that can be obtained
from our analysis of the models with smaller number of independent 
anomalous couplings introduced previously.
\begin{table}[t]
\centering
\caption{Limits on anomalous gauge boson couplings at the 95\% CL
for the models with three, two and one independent parameters.} \vskip 0.2cm
\begin{tabular}{|c|c|c|c|c|c|}
\hline
\multicolumn{6}{|c|}{Model with three
independent anomalous constants \cite{zeppenfeld1}:
$x_\gamma$, $y_\gamma$, $\delta_Z$;}\\
\multicolumn{6}{|c|}{
$x_Z=-\tan\theta_W\hskip 2pt x_\gamma$, $y_Z=\cot\theta_W\hskip 2pt
y_\gamma$.}\\ \hline
$\sqrt{s}\ (TeV)$ & $x_\gamma\hskip 2pt (10^{-3})$ & $\delta_Z\hskip 2pt
(10^{-3})$& $x_Z\hskip 2pt (10^{-3})$
& $y_\gamma\hskip 2pt (10^{-3})$& $y_Z\hskip 2pt (10^{-3})$
\\ \hline
$0.5$ & $-2.0\div 2.2$ & $-3.8\div 3.8$ & $-1.2\div 1.1$
&$-7.0\div 7.5$& $-12.8\div 13.7$
\\ \hline
$1  $ & $-0.6\div 0.6$ & $-1.1\div 1.1$ & $-0.3\div 0.3$
&$-4.0\div 4.5$ & $-7.3\div 8.2$
\\ \hline
\hline
\multicolumn{6}{|c|} {Model with
two independent anomalous constants \cite{zeppenfeld1}:  $x_\gamma$,
$y_\gamma$;}\\
\multicolumn{6}{|c|}{$\delta_Z=x_\gamma/2\sin\theta_W\cos\theta_W$,
$x_Z=-\tan\theta_W\hskip 2pt x_\gamma$, $y_Z=\cot\theta_W\hskip 2pt y_\gamma$.}
 \\ \hline
$\sqrt{s}\ (TeV)$ & $x_\gamma\hskip 2pt
(10^{-3})$ & $\delta_Z\hskip 2pt (10^{-3})$ & $x_Z\hskip 2pt(10^{-3})$
        & $y_\gamma\hskip 2pt (10^{-3})$ & $y_Z\hskip 2pt (10^{-3})$
\\ \hline
$0.5$ & $-1.8\div 1.8$ & $ -2.1\div 2.1 $ & $ -1.0\div 1.0$&$-6.6\div 6.8$&
$-12.1\div 12.4$    \\ \hline
$1$ & $-0.5\div 0.5$ & $ -0.6\div 0.6$ &$ -0.3\div 0.3$&$-3.0\div 2.4$&
$-5.5\div 4.4$\\ \hline
\hline
\multicolumn{6}{|c|}{
Model with one independent anomalous constant \cite{bilchak}:
$x_\gamma;$}\\
\multicolumn{6}{|c|}{
$x_Z=-\tan\theta_W$\hskip 2pt $x_\gamma=-\sin^2\theta_W$\hskip 2pt
$\delta_Z$.}\\
\hline
$\sqrt{s}\ (TeV)$ & $x_\gamma \hskip 2pt
(10^{-3})$ & $\delta_Z\hskip 2pt (10^{-3})$ & $x_Z\hskip 2pt(10^{-3})$
& $y_\gamma\hskip 2pt$ & $y_Z\hskip 2pt $
\\ \hline
$0.5$ & $-1.1\div 1.1$ & $  -2.6\div 2.6 $ &$ -0.6\div 0.6 $
& 0&0 \\ \hline
$1$ & $-0.3\div 0.3$ & $ -0.8 \div 0.8 $ &$-0.2  \div 0.2 $
&0&0 \\ \hline
\end{tabular}
\label{tab:tab2}
\end{table}
Comparing to the results in Tab.~\ref{tab:tab1}, one can observe that 
$\delta_Z$ can be more tightly constrained in this case than in the general one.
Concerning $y_\gamma$, the most stringent constraints are
obtained from the combination of $W_LW_L$ and $W_LW_T$ production channels.
In the case of the two-parameter model of Ref.~\cite{kuroda}, the bounds on
$x_\gamma$ and $\delta_Z$ are obtained in the same way as above, and are
numerically identical. Finally, in the two-parameter model of 
Ref.~\cite{zeppenfeld1}, due the relation (\ref{hisz}) among the couplings, 
$\sigma^L$ numerically proves to be more sensitive than
$\sigma^R$. Concerning final state polarizations, the bound on $x_\gamma$ is
obtained from $W_LW_L$ production, while that on $y_\gamma$ involves the
combination of both $LL$ and $TL+LT$ polarized cross sections.
\par
In summary, the obtained results indicate that the analysis of the 
cross sections of process (\ref{proc})  
with definite initial and final polarizations potentially allows to derive 
separate, and model dependent, contraints 
$CP$ conserving couplings in Eq.~(\ref{lagra}) with considerable sensitivity, 
typically of the order of $10^{-3}$ or better, depending on 
$E_{CM}$. Particularly stringent bounds can be expected for dynamical models 
beyond the SM with reduced number of independent couplings.\par


\newpage
\begin{thebibliography}{99}
\bibitem{gounaris1} G. Gounaris, D. Schildknecht and F. M. Renard, 
Phys.Lett. {\bf B263} (1991) 291.
\bibitem{anom2} A. A. Pankov and N. Paver, Phys.Lett. {\bf B324} (1994) 224.
\bibitem{hagiwara} K. Hagiwara, R. D. Peccei and D. Zeppenfeld, Nucl. Phys. 
{\bf B282} (1987) 253.
\bibitem{gounaris2} G. Gounaris, J. Layssac, G. Moultaka and F. M. Renard,
Int. J. Mod. Phys. A {\bf 8} (1993) 3285.
\bibitem{zeppenfeld1} K. Hagiwara, S. Ishihara, R. Szalapski and D. Zeppenfeld,
Phys. Lett. B {\bf 283}, 353 (1992); Phys. Rev. D {\bf 48}, 2182 (1993).
\bibitem{ruj} A. De Rujula, M.B. Gavela, P. Hernandez and E. Mass\'o,
Nucl. Phys. {\bf B384}, 3 (1992).
\bibitem{buch} W. Buchm\"uller and D. Wyler, Nucl. Phys. {\bf B268}, 621 (1986).
\bibitem{kuroda} M. Kuroda, J. Maalampi, K.H. Schwarzer and D. Schildknecht,
Nucl. Phys. {\bf B284}, 271 (1987).
\bibitem{bilchak} C. Bilchak, M. Kuroda and D. Schildknecht,
Nucl. Phys. {\bf B299}, 7 (1988).
\bibitem{frank} M. Frank, P. M\"attig, R. Settles and W. Zeuner, 
{\it ibidem}, p.223.
\bibitem{forty} R. W. Forty, J. B. Hansen, J. D. Hansen and R. Settles, 
{\it Collisions at 500 GeV: the Physics Potential}, Ed. P. M. Zerwas (1993), 
DESY 93-123C, p.235.
\bibitem{settles} R. Settles, {\it ibidem}, p. 591.
\bibitem{anlauf}  H. Anlauf, A. Himmler, P. Manakos, T. Mannel and H. Dahmen, 
Proceedings of the Workshop {\it Physics 
and Experiments with Linear $e^+e^-$ Colliders}, (Waikoloa, Hawaii, 1993), 
Eds. F. A. Harris, S. L. Olsen, S. Pakvasa and X. Tata, (World Scientific, 
Singapore 1993), p.708.
\bibitem{ruth} R.D. Ruth, Report SLAC-PUB-6751 (1995), to appear in the
{\it Report of the DPF Committee for Long Term Planning}.
\bibitem{anom4} V. V. Andreev, A. A. Pankov and N. Paver, Phys. Rev. D
{\bf 53} (1996) 2390.


\end{thebibliography}
\end{document}